\documentclass[10pt,twocolumn,aps,prb,floatfix,superscriptaddress]{revtex4-2}
\usepackage{graphicx}
\usepackage[dvipsnames]{xcolor}
\usepackage{amsmath}
\usepackage{amssymb}
\usepackage{comment}

\usepackage[colorlinks=true,
            linkcolor=blue,
            citecolor=blue,
            urlcolor=blue]{hyperref}
\begin{document}
\title{Magneto-optical properties of the neutral silicon-vacancy center in diamond under extreme isotropic strain fields}
\date{\today}

 \author{Meysam Mohseni}
 \affiliation{Wigner Research Centre for Physics, PO.\ Box 49, H-1525 Budapest, Hungary}
 \affiliation{E\"otv\"os Lor\'and University, P\'azm\'any P\'eter S\'et\'any 1/A, H-1117 Budapest, Hungary}

 \author{Gerg\H{o} Thiering}
\affiliation{Wigner Research Centre for Physics, PO.\ Box 49, H-1525 Budapest, Hungary} 
 
 \author{Adam Gali}
\affiliation{Wigner Research Centre for Physics, PO.\ Box 49, H-1525 Budapest, Hungary} 
\affiliation{Budapest University of Technology and Economics, Institute of Physics, Department of Atomics Physics, M\H{u}egyetem rkp.\ 3., 1111 Budapest, Hungary}

\begin{abstract}
The neutral silicon--vacancy (SiV$^{0}$) center in diamond combines inversion symmetry with optical emission, making it a robust quantum emitter resilient to stray electric fields. Using first-principles density-functional theory, we quantify its response to isotropic strain spanning strong compression and tensile regimes (effective hydrostatic pressures of approximately $-80$ to $180$~GPa). The coexistence of doubly degenerate $e_g$ and $e_u$ levels produces a structural instability captured by a quadratic product Jahn--Teller model. Under isotropic compression, the zero-phonon line blue-shifts nearly linearly while the $E_g$ phonon stiffens, suppressing vibronic instabilities and reducing Jahn--Teller quenching. Consequently, the Ham-reduced excited-state spin--orbit splitting increases substantially and the dark--bright vibronic gap widens. In contrast, isotropic tensile strain enhances vibronic effects and induces symmetry breaking beyond a critical strain, with tunneling-mediated dynamical averaging at the onset. Throughout the symmetry-preserving regime, parity remains well defined, so isotropic strain alone does not activate the dark transition. Charge-transition levels indicate photostability of the emission deep into the compressive regime, and near the highest photostable deformation ($\sim 100$~GPa), the radiative lifetime increases due to a reduced transition dipole moment despite the increasing optical energy. These trends yield compact calibration relations linking optical and spin observables to isotropic strain and establish SiV$^{0}$ as a symmetry-protected, strain-tunable quantum emitter operating into the multi-megabar-equivalent regime.
\end{abstract}

\maketitle

\section{Introduction}
\begin{figure*}
\includegraphics[width=\textwidth]{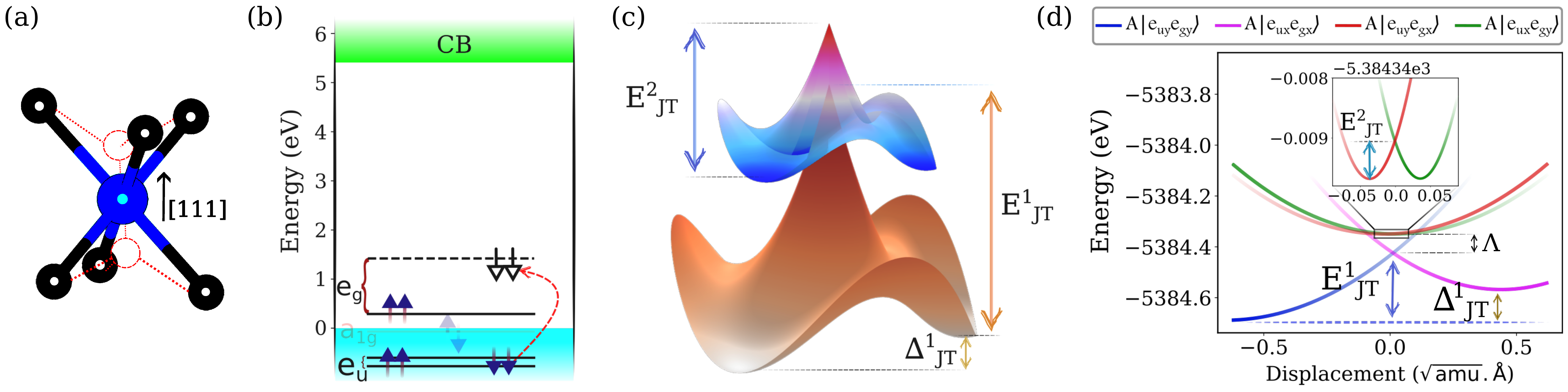}
\caption{
(a) Atomic configuration of the SiV defect in diamond, consisting of a single silicon atom positioned between two adjacent carbon vacancies and coordinated by six neighboring carbon atoms (illustrated in blue, red, and black, respectively).
(b) Electronic structure of the SiV triplet ground state. The red dashed arrow marks the electron excitation from the $e_u$ to the $e_g$ orbital, corresponding to the transition from $^3A_g$ to $^3E_u$.
(c) Adiabatic potential energy surface (APES) of the quadratic product Jahn--Teller (pJT) system along the ionic coordinates. The instability energies, $E^i_{JT}$, are indicated for constructive ($i=1$) and destructive ($i=2$) interference of the two orbital branches. Axial asymmetry, a second-order effect, is described by the parameter $\Delta^i_{JT}$.
(d) DFT-calculated potential energy surfaces along one-dimensional cuts. The $D_{3d}$ high-symmetry configuration (at 0.0~$\sqrt{\mathrm{amu}},\text{\AA}$) is unstable on two surfaces, in agreement with the pJT model. The splitting $\Lambda$ arises from static electronic correlation. Numerical values are provided in Table~\ref{tabale:PJT}.
}
\label{fig:levels}
\end{figure*}
The nitrogen–vacancy (NV) defect is the reference standard for quantum sensing in diamond~\cite{Rondin2014, Barry2020, Rembold2020, Schirhagl2014} and also applied for quantum communications~\cite{HansonNature2013, Hensen2015, Science2017, Science2021}. However, its polar \(C_{3v}\) symmetry leaves the coherent zero-phonon-line (ZPL) optical transition vulnerable to static and stray electric fields~\cite{Doherty2013,Rondin2014} which requires special tuning of the ZPL emission to produce indistinguishable single photon emitters which is a basic requirement for quantum communication~\cite{HansonNature2012}. 
This has motivated a shift toward inversion symmetric group-IV--vacancy centers (G4V), including SiV, GeV, SnV, PbV with $D_{3d}$ symmetry. These defects suppress the linear Stark effect and concentrate emission into a narrow, bright ZPL~\cite{Clark1995, Goss1996, Neu2011, Neu2013, Hepp2014, Rogers2014a, Sipahigil2014, Muller2014, Thiering2018, DeSantis2021}. The negatively charged G4V centers exhibit transform-limited lines and high Debye–Waller factors, yielding indistinguishable photons and all-optical spin control~\cite{Sipahigil2014, Sipahigil2016, Becker2016, Becker2018}. Together, first-principles studies and experimental measurements indicate that their optical and spin properties can remain robust under extreme compression~\cite{Vindolet2022, Razgulov2025, Mohseni2025}. These advantages position the G4V family as symmetry-protected alternatives to NV for photonic and sensing in environments where electric field noise and pressure are not negligible.

Broad implantation and growth studies have mapped the negatively charged G4V family and established reliable creation routes~\cite{Smith2019, Bradac2019, Iwasaki2015, Tchernij2017, RiedrichMoeller2014}. 
Diamond nanophotonics already integrates SiV centers into on-chip networks, including cavity-coupled devices and on-chip entanglement~\cite{Sipahigil2016, Nguyen2019, Pingault2014}. The same inversion symmetry that stabilizes the ZPL also suppresses the Stark effect—important for generating indistinguishable photons for long-range entanglement of remote emitters~\cite{DeSantis2021, Sipahigil2014, Sipahigil2016, Trusheim2018}. Pressure is therefore a natural control parameter, where theory and experiment predict a blue shift of negatively charged G4V ZPLs with increasing pressure, with larger deformation potentials for heavier group-IV species~\cite{Vindolet2022, Razgulov2025, Mohseni2025}. We now focus on the most studied silicon-vacancy (SiV) defect among the G4V centers.

Removing one electron yields a neutral SiV center (SiV$^0$) with an orbital triplet ($S=1$) ground state and millisecond spin coherence at 4.8 and 15~K temperatures~\cite{Rose2018}, a clear advantage over the negatively charged SiV center possessing the same coherence time at hundred millikelvin temperature regime~\cite{Sukachev2017}.  The emission is dominated by zero-phonon optical transition at 946~nm; however, the temperature dependence of photoluminescence and theoretical considerations invoking product Jahn--Teller theory reveal a dark \(^{3}A_{2u}\) level only $\sim7$~meV below the bright \(^{3}E_{u}\) level~\cite{Johansson2011, Rose2018, Thiering2019PJTE, Green2019PRB}.  Uniaxial stress mixes parity and activates a weak 951~nm line, confirming the doublet structure of the optical excited state~\cite{Green2019PRB}. Evidence of exceptionally long coherence in low-strain samples and nearly strain-free photophysics in nanofabricated resonators suggests the neutral center is a promising qubit and sensor~\cite{Rose2018, Rose2017, Zhang2020}. However, very little is known about magneto-optical properties of the neutral SiV center under extreme pressure.

In this study, we investigate the effects of hydrostatic compression on the $e_{g,u}$ orbitals by first principles simulations, which are responsible for both the optical transition and the zero-field splitting, while preserving the \(D_{3d}\) symmetry. By examining the pressure dependence of the 946~nm and 951~nm optical lines, as well as the spin Hamiltonian, we aim to benchmark the predicted deformation potentials, explore electron–phonon coupling within the product Jahn-Teller manifold, and assess the potential of SiV\(^{0}\) as a fiber-addressable pressure gauge for extreme environments. 

While hydrostatic compression provides a symmetry-preserving tuning knob relevant to diamond-anvil-cell conditions, recent experiments have demonstrated that diamond nanostructures can also sustain exceptionally large elastic tensile stresses without mechanical failure. In particular, reversible tensile strains approaching the theoretical elastic limit—up to $\sim 9\%$—have been achieved in nanoscale diamond needles and microfabricated diamond bridges, corresponding to tensile stresses on the order of $\sim 80$--$100$~GPa, with strain fields that are either localized (nanoneedles) or sample-wide and uniform (microbridges)~\cite{Banerjee2018, Dang2021}. Motivated by these results, we extend our study to the negative hydrostatic-pressure (isotropic tensile) regime and investigate uniform tensile strains up to $8\%$, corresponding to effective pressures of several tens of gigapascals, remaining within an experimentally realistic elastic window.

\section{Methodology}\label{Sec:Method}

All first-principles calculations were performed within Kohn–Sham density functional theory (DFT) using the projector–augmented–wave (PAW) formalism as implemented in \textsc{VASP}~\cite{Kresse1996_VASP_efficiency, Kresse1999_PAW_VASP, Blochl1994_PAW}. The neutral silicon–vacancy defect was modeled in a $4\times4\times4$ cubic diamond supercell (512 atoms), sampled at the $\Gamma$-point. A plane–wave kinetic–energy cutoff of 540~eV was used throughout. The screened hybrid functional HSE06 was employed for electronic structure, total–energy, magnetic properties, and defect–level calculations~\cite{Heyd2003_HSE, Krukau2006_HSE06}. All atomic positions were relaxed until the residual Hellmann–Feynman forces were below $10^{-2}$~eV/\AA. 

Excited-state geometries were obtained using the constrained $\Delta$SCF procedure by enforcing the target Kohn–Sham occupation for the defect manifold~\cite{Gali2009scf}. The ZPL energies were evaluated as the adiabatic energy difference between the fully relaxed minima of the ground and excited states on their respective adiabatic potential energy surfaces (APES),
\begin{equation}\label{eq:zpl}
E_{\mathrm{ZPL}} = E_{\mathrm{exc}}^{\min} - E_{\mathrm{gs}}^{\min}.
\end{equation}

Charge transition levels (CTLs) were computed using the Freysoldt–Neugebauer–Van~de~Walle (FNV) finite–size correction scheme~\cite{Freysoldt2009_FNV_PRL, Freysoldt2014_RMP}. Both electrostatic image–charge and potential–alignment terms were included,
\begin{equation}\label{eq:fnv}
E_{\mathrm{corr}}(q) = E_{\mathrm{el}} + q\,\Delta V ,
\end{equation}
and the thermodynamic transition levels were referenced to the valence band maximum (VBM) as
\begin{equation}\label{eq:CTL}
\varepsilon(q/q') = \frac{(E_{\mathrm{tot}}^{q}+E_{\mathrm{corr}}(q)) - (E_{\mathrm{tot}}^{q'}+E_{\mathrm{corr}}(q'))}{q'-q} - E_{\mathrm{VBM}}.
\end{equation}

\section{Results}

In this work, we investigate the response of the neutral SiV center to extreme isotropic strain by studying its magnetic and optical properties over the range $-80$ to $180$~GPa. For clarity, the term SiV refers to the neutral SiV optical center, and we omit the 'zero' when referring to its charge state in the context. We benchmark our first-principles results against the experimental and theoretical studies of Green \textit{et al.}~\cite{Green2017PRL,Green2019PRB} to assess whether the dark transition near $954$~nm can be activated under purely isotropic strain loading. We further evaluate the suitability of SiV as a quantum sensor under extreme conditions and establish calibration relations that link experimentally accessible spectroscopic observables to the applied pressure, enabling high-pressure quantum metrology.

\subsection{Atomic and Electronic Structure}\label{sec:atomic-electronic}
\begin{figure}
    \centering
    \includegraphics[width=1\linewidth]{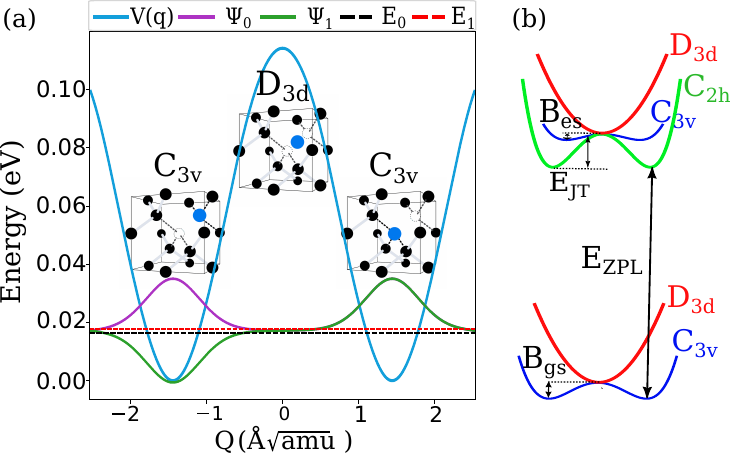}
    \caption{
    (a) Adiabatic double-well potential $V(Q)$ along the mass-weighted distortion coordinate
    $Q$ ($\text{\AA}\sqrt{\mathrm{amu}}$), illustrating tensile-strain--induced symmetry breaking from the
    high symmetry $D_{3d}$ configuration at $Q=0$ to two symmetry equivalent low symmetry $C_{3v}$
    minima at $Q=\pm Q_0$. The even-parity ground state $\psi_0$ and odd-parity first excited state $\psi_1$,
    with eigenenergies $E_0$ and $E_1$, are indicated by horizontal lines. Insets show the relaxed atomic
    structures at the $C_{3v}$ minima and at the $D_{3d}$ saddle point.
    (b) Schematic two-dimensional view of the adiabatic potential energy surface of SiV in the ground
    and excited states under $4\%$ tensile strain. The barrier energy $B_{\mathrm{gs/es}}$ denotes the energy
    difference between the $C_{3v}$ minima and the $D_{3d}$ saddle point in the ground/excited state,
    respectively.
    }
    \label{fig:DW}
\end{figure}
\begin{table}[!ht]
    \centering
    \caption{Tunneling rates $\nu_{\rm tun}$ between symmetry-equivalent $C_{3v}$ minima as a function of
    isotropic tensile strain, together with the corresponding effective hydrostatic pressures obtained
    from the equation of state.}
    \begin{ruledtabular}
    \begin{tabular}{cccc}
    Tensile (\%) & Pressure (GPa) & Barrier (meV)&$\nu_{\rm tun}$ (MHz) \\ \hline
    4.00 & $-50.263$ & 115 &22430.0 \\
    5.00 & $-58.159$ & 186 &88.7 \\
    6.00 & $-65.105$ & 274 &23.3 \\
    7.00 & $-70.941$ & 383 &2.8 \\
    8.00 & $-76.217$ & 513 &0.9 \\
    \end{tabular}
    \end{ruledtabular}
    \label{tab:tun}
\end{table}
\begin{table*}[ht]
    \centering
    \caption{ HSE06-calculated ZPL and pJT parameters under isotropic tensile from $\sim -50$ to 180~GPa. The Jahn--Teller stabilization energies $E_\text{JT}^i$ and phonon frequency $\hbar\omega$ are obtained by analyzing the potential energy surfaces shown in Fig.\ref{fig:levels}(c,d). The vibronic coupling constants $F_g$ and $F_u$ are evaluated using Eqs.\ref{eq:fg,u}. The parameter $\delta$ denotes the vibronic energy difference between the $^3A_{2u}$ and $^3E_{u}$ states. The quantities $\Lambda$ and $\Xi$ are extracted by solving Eq.\ref{eq:e-correction-Hamiltonian}.
}
\begin{ruledtabular}
    \begin{tabular}{cccccccccc}
        Strain (GPa) & $E_\text{ZPL}$ (eV) & $E_\text{JT}^1$~(meV) & $E_\text{JT}^2$~(meV) & $\hbar\omega$ (meV) & $\Lambda$ (meV) & $\delta$ (meV) & $\Xi$ (meV) & $F_g$ & $F_u$ \\ \hline
        $-50.09$ & 1.96 & $-$ & $-$ & $-$ & $-$ & $-$ & $-$ & $-$ & $-$ \\
        $-41.20$ & 1.30 & 285.8 & 0.91 & 71.99 & 86.1 & 5.24 & 0.46 & 107.15 & 95.70 \\
        0.00 & 1.35 (1.31\footnotemark[1]) & 257.32 & 0.45 & 77.39 & 81.94 & 6.62 (6.8\footnotemark[2]) & 52.52 & 103.96 & 95.61 \\ 
        27.10 & 1.41 & 238.39 & 0.25 & 78.94 & 80.67 & 7.28 & 52.71 & 100.15 & 93.86 \\ 
        67.70 & 1.44 & 223.73 & 0.10 & 79.44 & 80.55 & 7.68 & 53.22 & 96.26 & 92.27 \\ 
        117.80 & 1.47 & 212.17 & 0.03 & 79.93 & 80.41 & 8.00 & 53.69 & 93.18 & 90.99 \\ 
        180.00 & 1.49 & 209.07 & 0.00 & 81.32 & 80.27 & 8.28 & 54.03 & 92.20 & 92.20 \\ 
    \end{tabular}
    \footnotetext[1]{Ref.~\cite{Green2019PRB}}
    \footnotetext[2]{Ref.~\cite{Green2017PRL}}
\end{ruledtabular}
\label{tabale:PJT}
\end{table*}
The silicon vacancy in diamond adopts a split-vacancy configuration, where the Si atom relaxes into the bond center between two missing carbon sites. This arrangement preserves inversion symmetry and gives the defect $D_{3d}$ point group symmetry, as illustrated in Fig.~\ref{fig:levels}(a). The dangling bonds from the six nearest-neighbor carbons form localized orbitals of $e_u$ and $e_g$ symmetry, positioned near the valence band maximum and within the band gap, respectively, depicted in Fig.~\ref{fig:levels}(b).  

In the neutral charge state, the $e_u$ orbital is fully filled with four electrons, while the $e_g$ orbital is half filled with two electrons. The occupancy of the double-degenerate $e_g$ level leads to a high-spin ground state, stabilized as a spin triplet ($S=1$). The corresponding total wavefunction transforms as ${}^3A_{2g}$ under $D_{3d}$ symmetry, with the symmetric spin component paired to an antisymmetric orbital function, as required by the Pauli principle. In this picture, the ground-state configuration is $e_g^2 = e_{gx}^1 e_{gy}^1$, and the orbital part of the wavefunction is explicitly written as
\begin{equation}
A(e_{gx} e_{gy}) = \frac{1}{\sqrt{2}} \Big[ e_{gx}(r_1) e_{gy}(r_2) - e_{gx}(r_2) e_{gy}(r_1) \Big],
\end{equation}
which captures the antisymmetry of the spatial component and underpins the $^3A_{2g}$ ground state~\cite{ciccarino2020}.  

Optical excitation promotes one electron from the $e_u$ to the $e_g$ orbital, resulting in an $e_g^1 e_u^1$ configuration. In this case, four distinct antisymmetrized orbital wavefunctions can be constructed:
\begin{align}
A(e_{ux} e_{gx}), \quad
A(e_{uy} e_{gx}), \quad
A(e_{ux} e_{gy}), \quad
A(e_{uy} e_{gy}),
\end{align}
which form the basis of the excited-state manifold. Linear combinations of these states decompose into the irreducible representations ${}^3E_u$ and ${}^3A_{2u}$, defining the symmetry-adapted excited triplet states of the defect. These excited states are Jahn--Teller active and couple strongly to lattice vibrations of $E_g$ symmetry, leading to vibronic instabilities that govern much of the defect’s optical and spin--orbit behavior.

We now turn to the negative effective hydrostatic-pressure (isotropic tensile) regime. In contrast to compression, tensile strain progressively softens the relevant vibrational modes and can induce symmetry breaking in the adiabatic landscape. For small tensile strains, $0$--$3\%$ ($0$ to $\sim-41$~GPa in our equation of state), the system remains in the high-symmetry $D_{3d}$ regime. In this interval, structural relaxation does not produce a static distortion, and the defect retains inversion symmetry.

Beyond $\sim4\%$ tensile strain, structural relaxation yields two symmetry-equivalent low-symmetry configurations of $C_{3v}$ symmetry~\cite{Yue2026}. Since these two minima are related by symmetry, the distortion coordinate
$Q$ can be chosen along the minimum-energy path such that the adiabatic landscape is well approximated by a symmetric double-well potential, $V(Q)=V(-Q)$, with minima at $Q=\pm Q_0$ separated by a barrier at the high-symmetry configuration $Q=0$. This tensile-strain--induced symmetry breaking and the corresponding adiabatic potential energy surfaces are illustrated in Fig.~\ref{fig:DW}. In this situation, the low-energy nuclear eigenstates are not strictly localized in a single minimum, while they are tunneling through the barrier hybridizes left- and right-localized configurations into even- and odd-parity combinations, forming a near-degenerate doublet~\cite{LandauLifshitzQM,LeggettRMP1987}.

To quantify the interconversion between the two equivalent $C_{3v}$ configurations, we construct a one-dimensional effective nuclear Hamiltonian along $Q$,
\begin{equation}
\hat H_{\rm nuc} = -\frac{\hbar^2}{2M}\frac{d^2}{dQ^2} + V(Q),
\end{equation}
where $V(Q)$ is obtained from the calculated energy profile along the distortion path and $M$ is the effective mass associated with the collective coordinate, defined by the mass-weighted atomic displacements.
Solving the corresponding one-dimensional Schr\"odinger equation yields the lowest eigenenergies $E_0$ and $E_1$, which for a symmetric double well correspond to the even-parity ground state and odd-parity first excited state. The tunneling splitting is defined as
\begin{equation}
\Delta E = E_1 - E_0,
\end{equation}
and the associated tunneling (interconversion) rate is
\begin{equation}
\nu_{\rm tun} = \frac{\Delta E}{h}.
\end{equation}
In the semiclassical regime, $\Delta E$ may alternatively be estimated using WKB methods, yielding an exponentially small splitting controlled by the under-barrier action~\cite{LandauLifshitzQM,GargAmJPhys2000}. At $4\%$ tensile strain with a barrier energy of $V(0)=115$~meV, we obtain a large tunneling rate of $\nu_{\rm tun}\approx22.43$~GHz, implying rapid
dynamical averaging such that the defect effectively retains $D_{3d}$ symmetry on typical steady-state experimental timescales. With increasing tensile strain, the barrier height grows and tunneling is
progressively suppressed, dropping into the MHz regime, as summarized in Table~\ref{tab:tun}.

In the symmetry-broken tensile regime ($\gtrsim4\%$), the electronic structure is naturally described within the $C_{3v}$ point group. The ground state remains a spin triplet ($S=1$): one $e$-type defect level, denoted $e_1$, is fully occupied, while a higher $e$-type level ($e_2$) is half occupied with two electrons distributed over its doublet components. The resulting $e_2^2$ configuration yields an orbital singlet with triplet spin, transforming as ${}^3A_{2}$ in $C_{3v}$. Analogous to the $D_{3d}$ case, the orbital part of the antisymmetrized two-electron wavefunction in the half-filled $e_2$ manifold can be written as
\begin{equation}
A(e_{2x} e_{2y}) = \frac{1}{\sqrt{2}}
\Big[ e_{2x}(r_1) e_{2y}(r_2) - e_{2x}(r_2) e_{2y}(r_1) \Big],
\end{equation}
which captures the antisymmetry of the spatial component and underpins the ${}^3A_{2}$ ground
state.

Optical excitation promotes an electron from the fully occupied $e_1$ level to the half-filled $e_2$ level, producing an $e_1^1 e_2^1$ configuration. In this basis, four antisymmetrized orbital product functions can be constructed,
\begin{align}
A(e_{1x} e_{2x}), \quad
A(e_{1y} e_{2x}), \quad
A(e_{1x} e_{2y}), \quad
A(e_{1y} e_{2y}),
\end{align}
which span the excited-state orbital manifold. Symmetry-adapted linear combinations decompose into the irreducible representations ${}^3E$ and ${}^3A_{2}$ of $C_{3v}$, defining the tensile-regime excited triplet states. These states remain Jahn--Teller active and couple strongly to vibrational modes of $E$ symmetry, preserving the same core vibronic mechanism that governs the optical and spin--orbit response of the SiV center.

\subsection{Jahn--Teller Interaction and Spin--Orbit Coupling}\label{sec:JT}

The lowest optically addressable excited configuration of SiV in $D_{3d}$ configuration is $e_g^1 e_u^1$, which is unstable to symmetry-lowering distortions and couples to the quasi-localized $E_g$ phonon doublet. This realizes a pJT system of type $E_g \otimes e_u \otimes e_g$,
in which both electronic doublets ($e_u$ and $e_g$) are Jahn--Teller active and interact with the same $E_g$ vibrational mode~\cite{Thiering2019PJTE,ciccarino2020}. Within a first-order (linear) vibronic treatment, the coupled spin--vibronic dynamics in the triplet manifold is captured by the total Hamiltonian
\begin{align}
\hat{H} = \hat{H}_{\mathrm{osc}} + \hat{H}_{\mathrm{pJT}} + \hat{W} + \hat{H}_{\mathrm{SO}} .
\end{align}

The $E_g$ vibrational doublet is modeled as a two-dimensional harmonic oscillator
$\hat{H}_{\mathrm{osc}} = \hbar \omega_E \sum_{i = x, y} \left( a_i^\dagger a_i + \tfrac{1}{2} \right)$
with creation and annihilation operators $a_i^\dagger$, $a_i$, and  $\hbar \omega_E$ as the phonon frequency (listed in Table~\ref{tabale:PJT}).

The linear vibronic interaction in the $e_u \otimes e_g$ orbital subspace is
\begin{align}
\hat{H}_{\mathrm{pJT}} &=
F_u \big( \hat{X}\,\sigma_z \otimes \sigma_0 + \hat{Y}\,\sigma_x \otimes \sigma_0 \big) \nonumber \\
&+ F_g \big( \hat{X}\,\sigma_0 \otimes \sigma_z + \hat{Y}\,\sigma_0 \otimes \sigma_x \big) ,
\end{align}
where $\sigma_{x,z}$ are Pauli matrices acting on the $e_{u}$ and $e_{g}$ orbital doublets, $\hat{X} = \frac{a_x^\dagger + a_x}{\sqrt{2}}$ and $\hat{Y} = \frac{a_y^\dagger + a_y}{\sqrt{2}}$ and $\sigma_{0}$ is the two-dimensional unit matrix that is introduced for the individual electron–phonon coupling strength $F_{u/g}$~\cite{Thiering2019PJTE}. The stabilization energies of the two adiabatic branches follow from
\begin{align}\label{eq:fg,u}
E_{\mathrm{JT}}^{(1)} &= \frac{(F_g + F_u)^2}{2\hbar\omega_E}, \\
E_{\mathrm{JT}}^{(2)} &= \frac{(F_g - F_u)^2}{2\hbar\omega_E} .
\end{align}

At ambient conditions, we obtain $E_{\mathrm{JT}}^{(1)} = 257.32$~meV and $E_{\mathrm{JT}}^{(2)} = 0.45$~meV with $\hbar\omega_E = 77.39$~meV. Under compression to $180$~GPa, $E_{\mathrm{JT}}^{(1)}$ decreases to $\sim 209$~meV, while $E_{\mathrm{JT}}^{(2)}$ nearly vanishes. This indicates that pressure suppresses vibronic instabilities, stabilizing the high-symmetry $D_{3d}$ geometry. The increase of $\hbar\omega_E$ from 77.4 to 81.3~meV reflects stiffening of the lattice vibrations. At the tensile stability limit of the $D_{3d}$ configuration (3\% tensile strain, $\sim-41$~GPa), $E_{\mathrm{JT}}^{(1)}$ increases to $285.8$~meV while $E_{\mathrm{JT}}^{(2)}$ remains of the same order ($0.91$~meV), accompanied by a softening of the $E_g$ phonon to $\hbar\omega_E = 71.99$~meV (Table~\ref{tabale:PJT}).

The electronic splittings among the $^3A_{1u}$, $^3A_{2u}$, and $^3E_{u}$ triplets are described by~\cite{Thiering2019PJTE,ciccarino2020}
\begin{align}\label{eq:e-correction-Hamiltonian}
\hat{W} &= 
\Lambda \Big( |{}^3A_{1u}\rangle\langle{}^3A_{1u}| - |{}^3A_{2u}\rangle\langle{}^3A_{2u}| \Big) \nonumber \\
&- \Xi \Big( |{}^3E_{ux}\rangle\langle{}^3E_{ux}| + |{}^3E_{uy}\rangle\langle{}^3E_{uy}| \Big),
\end{align}
with parameters $\Lambda$ and $\Xi$ taken from Table~\ref{tabale:PJT}. At $0$~GPa, $\Lambda = 81.9$~meV and $\Xi \approx 52.5$~meV. These values remain nearly constant under pressure, indicating that static electronic correlations are less sensitive to compression than the vibronic couplings. At 3\% tensile strain ($\sim-41$~GPa), we obtain $\Lambda = 86.1$~meV and $\Xi = 0.46$~meV, indicating that while $\Lambda$ remains comparable to its ambient-pressure value, $\Xi$ is strongly reduced as tensile strain approaches the structural instability.

In the excited manifold, the spin--orbit (SO) interaction is
\begin{align}
\hat{H}_{\mathrm{SO}} &= m_s \Bigg[
\frac{\lambda_u}{2}\,(\sigma_y \otimes \sigma_0) 
+ \frac{\lambda_g}{2}\,(\sigma_0 \otimes \sigma_y) 
\Bigg],
\end{align}
where $m_s\in\{+1,0,-1\}$ is the spin projection and $\lambda_{u/g}$ are the single-hole SO parameters on $e_{u/g}$~\cite{ciccarino2020}. In the strong vibronic regime, the observable splitting is Ham-reduced to~\cite{Ham1965}
\begin{align}
\Delta_{\mathrm{SO}} = p_u \lambda_u + p_g \lambda_g,
\end{align}
where the Ham reduction factors $p_u$ and $p_g$ are equivalent; therefore, $\Delta_{\mathrm{SO}} = p (\lambda_u + \lambda_g)$.
At $0$~GPa, $\Delta_{\mathrm{SO}} = 28.61$~GHz, rising monotonically to $\sim 74.7$~GHz at $180$~GPa (Table~\ref{tabale:lambda}). This enhancement reflects the reduction of vibronic quenching under pressure, as the JT stabilization energy decreases. In contrast to the compressive regime, at 3\% tensile strain, the Ham-reduced spin--orbit splitting is strongly suppressed, with $\Delta_{\mathrm{SO}} = 8.24$~GHz, consistent with enhanced vibronic quenching as the Jahn--Teller stabilization energy increases under tensile strain.

\begin{table}[h]
\centering
\caption{Calculated isotropic strain dependence of the Ham reduction factor ($p_{u,g}$) and the vibronic couplings in the excited state. The Ham-reduced spin-orbit (SO) splitting parameters $p_u\lambda_{u}$, $p_g\lambda_{g}$ and the resulting full spin-orbit splitting parameter $\Delta_{\mathrm{SO}}$ are listed.}
\begin{ruledtabular}
\begin{tabular}{ccccc}
\footnotesize{Strain (GPa)} & \footnotesize{$p_{u,g}$} & \footnotesize{$p_u\lambda_{u}$~(GHz)} & \footnotesize{$p_g\lambda_{g}$~(GHz)} & \footnotesize{$\Delta_{\mathrm{SO}}$~(GHz)} \\ \hline
$-41.20$   & 0.004 & 7.14  &1.10 & 8.24 \\
0.00   & 0.012 & 22.77 & 2.25 & 25.02 \\
27.10  & 0.016 & 31.34 & 6.52 & 37.86 \\
67.70  & 0.022 & 41.48 & 9.56 & 51.05 \\
117.80 & 0.025 & 47.67 & 12.16 & 59.83 \\
180.00 & 0.027 & 53.14 & 15.00 & 68.14 \\
\end{tabular}
\end{ruledtabular}
\end{table}\label{tabale:lambda}

The pJT interaction produces two low-lying vibronic states: a dark $^3\tilde{A}_{2u}$ state and an optically bright $^3\tilde{E}_u$ state~\cite{Thiering2019PJTE}. Their energy separation,
$\delta = E(^3\tilde{E}_u) - E(^3\tilde{A}_{2u})$, is $6.62$~meV at $0$~GPa, consistent with the experimentally observed $946/951$~nm doublet in which the dark state becomes optically active under uniaxial
stress~\cite{Green2019PRB,Green2017PRL}. Under hydrostatic compression, $\delta$ increases gradually to $8.28$~meV at $180$~GPa (Table~\ref{tabale:PJT}), reflecting phonon stiffening and the concomitant reduction of dynamic Jahn--Teller fluctuations. In contrast, at $3\%$ tensile strain ($\sim-41$~GPa), the vibronic splitting is reduced to $\delta = 5.24$~meV (Table~\ref{tabale:PJT}), consistent with enhanced dynamic Jahn--Teller fluctuations associated with phonon softening under tensile strain. Together, these results demonstrate that isotropic strain in the range from $-41.2$ to $180$~GPa does not activate the dark transition, which requires symmetry-breaking (non-hydrostatic) stress.

At higher tensile strains beyond $\sim4\%$, the inversion symmetry is broken and the system enters a symmetry-lowered regime. In this range, the $C_{3v}$ configurations remain energetically tied to the
high-symmetry $D_{3d}$ saddle point by a relatively small barrier of approximately $5$--$10$~meV between $4\%$ and $8\%$ tensile strain. In this regime, one might anticipate the emergence of an
$E \otimes e \otimes e$ product Jahn--Teller problem within the $C_{3v}$ point group, which would further lower the symmetry to $C_{1h}$. We did not analyze this excited state because we show below that the photoionization threshold energies are lower than the ZPL energies under this condition which prevents the realization of the pJT phenomena beyond $\sim4\%$ tensile strain.

\subsection{Optical Properties and Charge Transition Levels}\label{sec:optics-ctl}
\begin{figure}
\includegraphics[scale=.36]{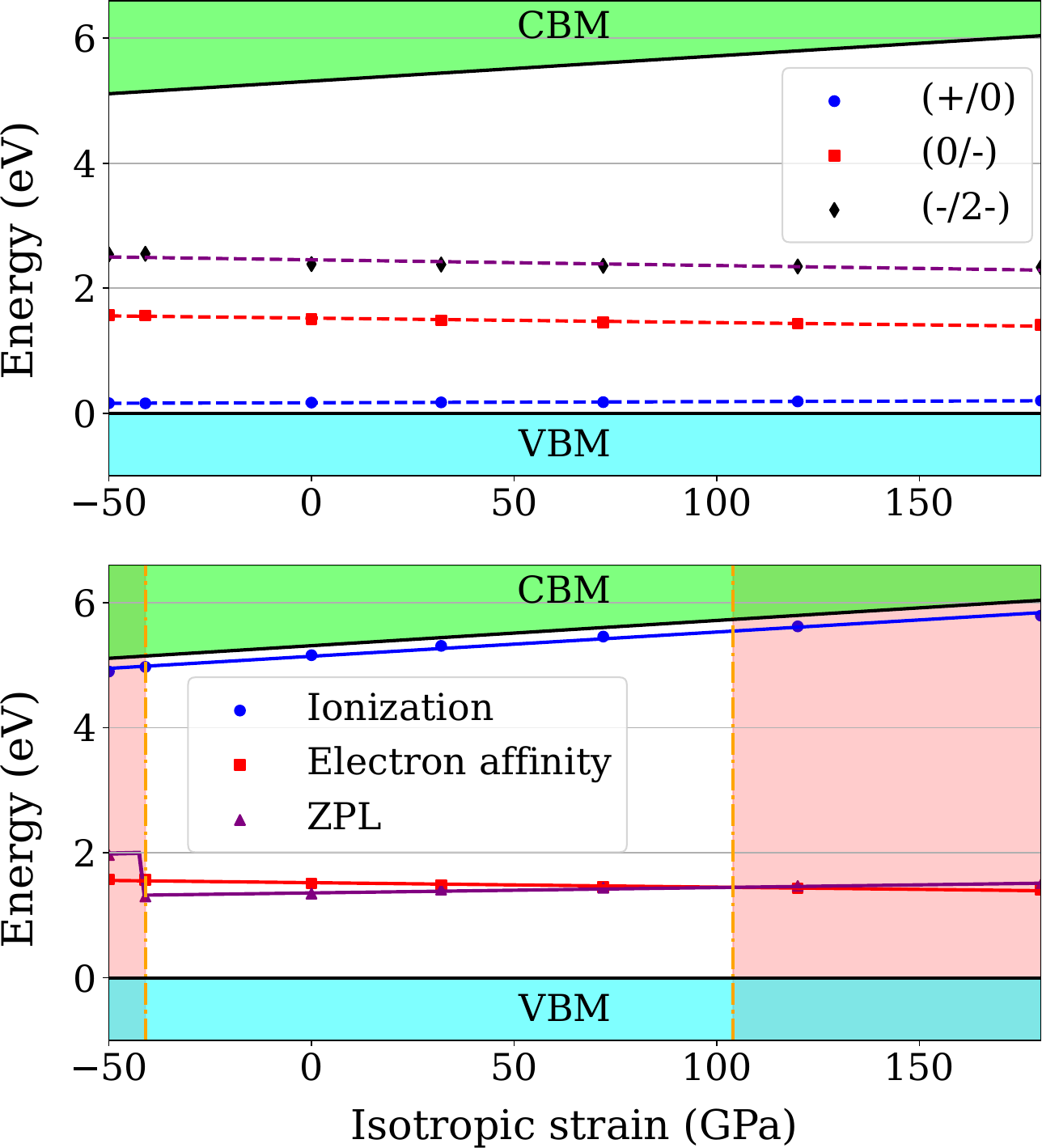}
\caption{
(a) Calculated charge transition levels of SiV defect under isotropic strain in the range of $-50$
to 180~GPa as obtained by HSE06 functional. (b) The ZPL shift was calculated by HSE06 functional. We highlight the strain regions by light red color where SiV is not photostable.
}
\label{fig:chg}
\end{figure}
The ZPL at each strain value was computed using Eq.~\ref{eq:zpl} by enforcing the $e_{u/1}\rightarrow e_{g/2}$ transition in the relaxed $D_{3d}$ and $C_{3v}$
configurations. At ambient conditions we obtain $E_{\mathrm{ZPL}}=1.35$~eV (experiment $1.31$~eV $=$ 946~nm in Ref.~\cite{Green2019PRB,Johansson2011}). In the symmetry-preserving $D_{3d}$ regime
corresponding to isotropic strains from $-41.2$ to 180~GPa, the ZPL blue-shifts nearly linearly from $1.30$~eV to $1.49$~eV, with an effective slope of about
$0.86$~meV/GPa. Within this regime we find no strain-induced reordering of the lowest vibronic levels. The optically bright ${}^3\tilde{E}_{u}$ remains above the
dark ${}^3\tilde{A}_{2u}$, and the vibronic gap $\delta$ evolves smoothly (Table~\ref{tabale:PJT}). Beyond about $4\%$ isotropic tensile strain the defect
relaxes into a symmetry-broken $C_{3v}$ configuration. This structural transformation leads to a sudden jump of the ZPL in Fig.~\ref{fig:chg}(b) and
coincides with the loss of photostability.

Charge stability was assessed from thermodynamic charge transition levels $\varepsilon(q/q')$ computed using Eqs.~\ref{eq:fnv} and \ref{eq:CTL}. Figure~\ref{fig:chg}(a) shows the evolution of the relevant CTLs under isotropic strain, while Fig.~\ref{fig:chg}(b) overlays the ZPL energy with the strain-dependent electron affinity and ionization energy, which define the lowest thresholds for carrier exchange with the bands. The neutral SiV is photostable under photoexcitation when $E_{\mathrm{ZPL}}$ lies below the lowest ionization threshold. Under isotropic compression the blue-shifting ZPL approaches and crosses the valence-band ionization threshold at about 100~GPa, marking the onset of photo-instability region where photoexcitation can capture a valence electron and convert the center to the negative charge state $\mathrm{SiV}^{0}{+}e^{-}\rightarrow\mathrm{SiV}^{-}$. In the tensile direction, the ZPL jump associated with the $C_{3v}$ regime drives $E_{\mathrm{ZPL}}$ above the electron-affinity line in Fig.~\ref{fig:chg}(b). In this strain range the defect is no longer photostable under illumination. The light-red regions in Fig.~\ref{fig:chg}(b) indicate the isotropic strain intervals where charge
conversion becomes energetically allowed.

\subsection{Hyperfine Parameters}
\begin{figure*}
\centering
\includegraphics[width=\textwidth]{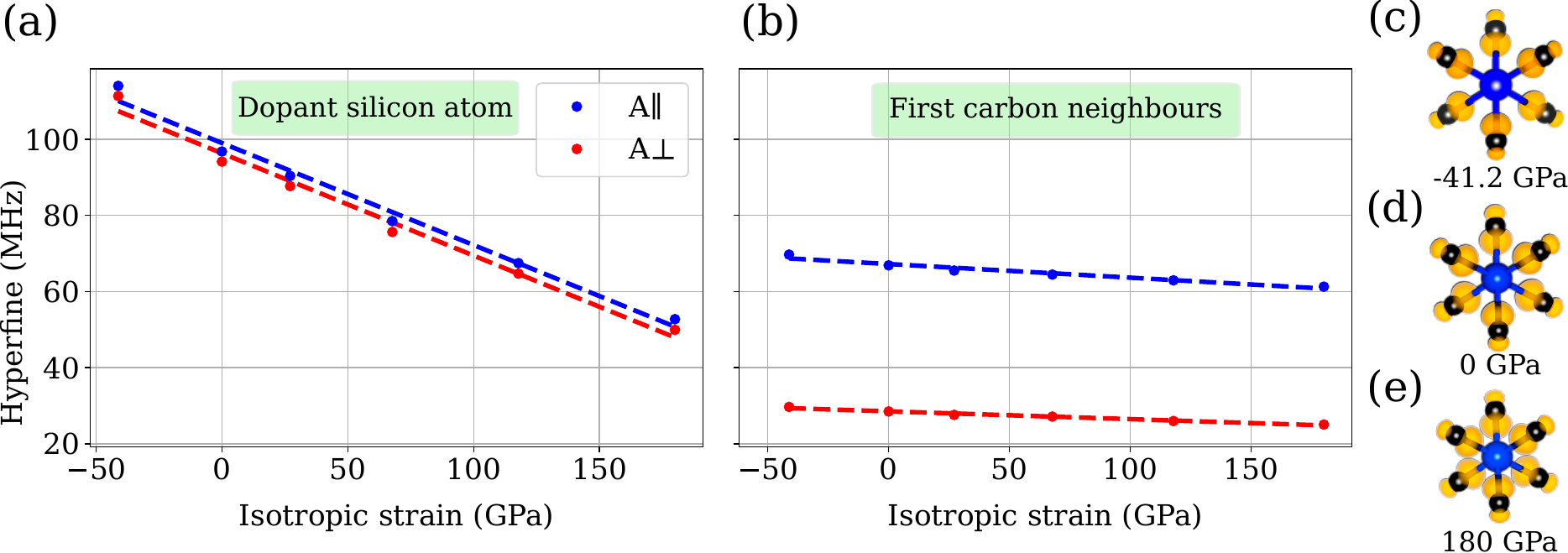}
\caption{
Hyperfine parameters of the neutral SiV defect within the region of isotropic strain resulting in high $D_{3d}$ symmetry, showing $A_\parallel$ and $A_\perp$ in the ground state for (a) the dopant $^{29}$Si atom and (b) the first $^{13}$C neighbor atoms. The ground-state spin density is illustrated for (c) $-41.2$~GPa, (d) zero and (e) 180~GPa pressures. The spin density is defined as the charge-density difference between the majority and minority spin channels in spin-polarized calculations; the isosurface corresponds to positive spin density, with an absolute value of $0.02~\text{\AA}^{-3}$. 
}
\label{fig:hf}
\end{figure*}
For the EPR-active triplet ground state of the SiV defect, we calculate the hyperfine parameters using the HSE06 hybrid functional. The hyperfine tensor describing the interaction between a nuclear spin at position $R_I$ and the defect spin density $\rho_s(r)$ is given by  
\begin{equation}
A_{ij}=\frac{4\pi}{3}\frac{g_N\gamma_N g\gamma_e}{\langle \hat S_z \rangle} 
\int \mathrm{d}^3r\, \rho_s(r)\, m_{ij}(r-R_I),
\end{equation}
where $g_N$ and $\gamma_N$ ($g$ and $\gamma_e$) denote the nuclear (electron) $g$ factors and gyromagnetic ratios, respectively. The interaction kernel  
$m_{ij}=\delta_{ij}\delta(r)-\tfrac{1}{2}(3x_ix_j-r^2\delta_{ij})r^{-5}$  
contains the Fermi-contact and dipole–dipole contributions.  Hyperfine tensors were evaluated for the central $^{29}$Si nucleus ($I{=}1/2$) and the nearest $^{13}$C nuclei using the spin density from HSE06. Restricting to the axial form appropriate for the defect symmetry, the Hamiltonian is written as~\cite{Felton2009}
\begin{equation}\label{eq:hf-hamiltonian}
\hat{H}_{\mathrm{HF}} = A_{\parallel} S_z I_z + A_{\perp}\,(S_x I_x + S_y I_y).
\end{equation}
When off–diagonal elements vanish by symmetry at the central site, the HF principal $z$-axis coincides with the defect symmetry axis. For general neighbors, the tilt of the HF principal axis is given by the polar angle $\theta$ from the diagonalization of the HF tensor. Core–polarization effects in the Fermi-contact term were included following Refs.~\cite{VandeWalle1993,Yazyev2005}.

Figure~\ref{fig:hf} summarizes our results for the hyperfine parameters of the dopant $^{29}\mathrm{Si}$ and its nearest-neighbor $^{13}\mathrm{C}$ atoms in the
ground state within the $D_{3d}$ regime spanning isotropic strains from $-41.2$ to 180~GPa. At ambient pressure, $^{29}\mathrm{Si}$ exhibits $A_{\parallel}=96.8$~MHz, $A_{\perp}=94.1$~MHz, and $\theta=0^\circ$, compared with the experimental values of 76.3, 78.9, and $0^\circ$, respectively \cite{Edmonds2008}. For the nearest-neighbor $^{13}\mathrm{C}$ atoms, our
calculations yield $A_{\parallel}=66.9$~MHz, $A_{\perp}=28.5$~MHz, and $\theta=34.6^\circ$, in good correspondence with the experimental values of
66.2, 30.2, and $35.3^\circ$ \cite{Edmonds2008}. Under isotropic compression, both $^{29}\mathrm{Si}$ and $^{13}\mathrm{C}$ hyperfine couplings decrease monotonically with increasing pressure. This
reduction arises from the progressive delocalization and spatial expansion of the spin density away from the Si and C ions as the lattice is compressed.

At $3\%$ isotropic tensile strain, corresponding to $-41.2$~GPa and still within the $D_{3d}$ symmetry regime, the hyperfine couplings increase relative to
ambient conditions. For $^{29}\mathrm{Si}$ we obtain $A_{\parallel}=114.01$~MHz, $A_{\perp}=111.36$~MHz, with $\theta=0^\circ$.
For the nearest-neighbor $^{13}\mathrm{C}$ atoms, the corresponding values are $A_{\parallel}=69.69$~MHz, $A_{\perp}=29.68$~MHz, and $\theta=34.73^\circ$.
These trends indicate an enhanced localization of the ground-state spin density on the defect core under tensile strain. 

It is noted that the HSE06 result on the magnitude of the hyperfine constants of $^{29}$Si apparently deviates from the experimental data at zero pressure but we expect to obtain an accurate trend for the strain dependence. According to our calculations in the isotropic strain region conserving the $D_{3d}$ symmetry, the variation in the hyperfine constants of $^{29}$Si is sublinear as shown in Fig.~\ref{fig:hf}(a). Nevertheless, we also fit our data points to a linear function, in order to make a direct comparison to a recent study that ran parallel to ours~\cite{Yue2026}. The obtained slope of Silicon $A_{\perp}\simeq A_{\parallel} =-5.58$~MHz/strain value follows the value of $-4.98$~MHz/strain achieved by r2SCAN functional~\cite{Yue2026} in that study which produces well the magnitude of hyperfine constants at zero pressure for $^{29}$Si~\cite{Yue2026}. Furthermore, the first carbon neighbors exhibit $A_{\perp}$ and $A_{\parallel}$ values of 0.76 and 0.42~MHz/strain, respectively, as calculated using the HSE06 hybrid functional, compared with 0.92 and 0.69~MHz/strain obtained using the r2SCAN functional~\cite{Yue2026}. We assume that HSE06 values are accurate for $^{13}$C hyperfine tensors.

\begin{figure*}
\centering
\includegraphics[width=\textwidth]{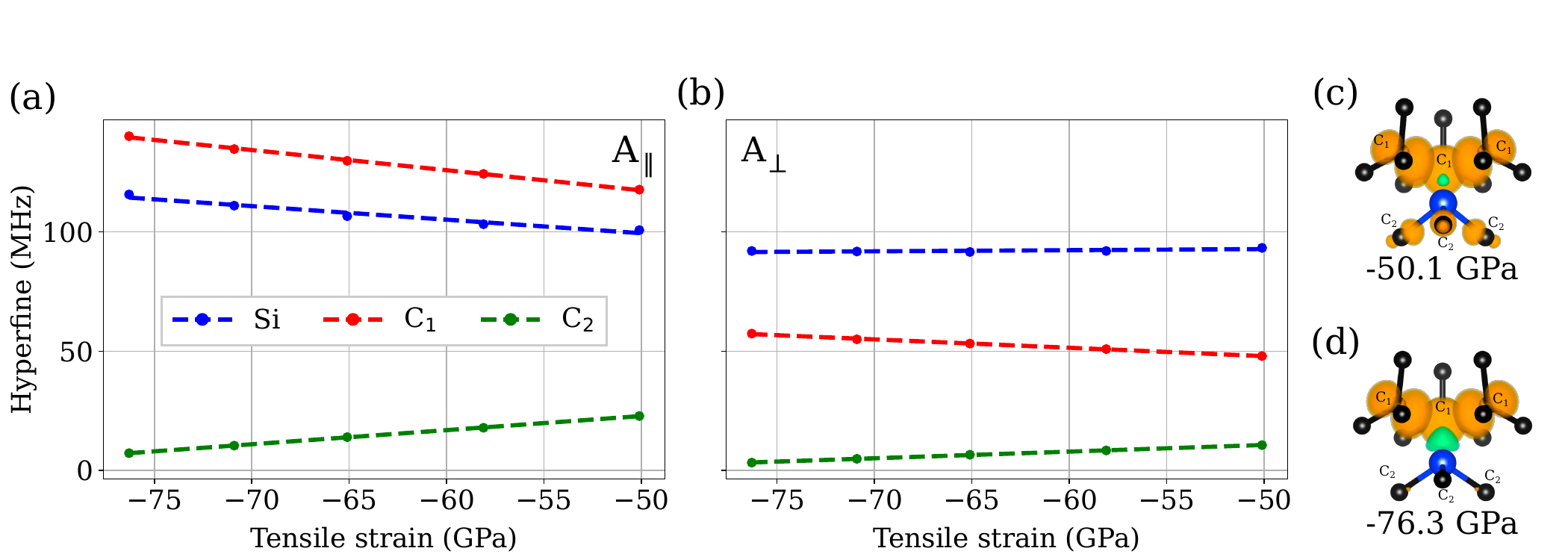}
\caption{
Hyperfine parameters of the neutral SiV defect within the region of isotropic tensile strain resulting in low $C_{3v}$ symmetry, showing $A_\parallel$ and $A_\perp$ in the ground state for (a) the dopant $^{29}$Si atom and (b) the first $^{13}$C neighbor atoms that have two groups. The ground-state spin density is illustrated for (c) $-50.1$~GPa and (d) $-77.6$~GPa tensile strain. The spin density is defined as the charge-density difference between the majority and minority spin channels in spin-polarized calculations; the isosurface corresponds to positive (orange lobes) and negative (green lobes) spin density, with an absolute value of $0.006~\text{\AA}^{-3}$. The linear fitting in dashed lines is used to guide the eyes.}
\label{fig:hfc3v}
\end{figure*}

At high tensile isotropic strain, the structure reconstructs to $C_{3v}$ symmetry which strongly affects the hyperfine constants of $^{29}$Si and the surrounding $^{13}$C nuclear spins. There will be two groups of $3\times$ $^{13}$C isotopes ($C_1$ and $C_2$) and the closely isotropic hyperfine field around $^{29}$Si nuclear spin breaks too. The calculated hyperfine constants are depicted in Fig.~\ref{fig:hfc3v}.

\subsection{Radiative lifetime}
Next, we computed the radiative lifetime ($\tau_{\rm rad}$) using the standard dipole emission formula for a dielectric medium, which is appropriate for complex point defects:
\begin{equation}
\label{eq:tau}
\Gamma_{\rm rad} = \frac{n\, \omega^3\,\mu^2}{3\pi\varepsilon_0\hbar c^3}\text{,} 
\quad \tau_{\rm rad} = \frac{1}{\Gamma_{\rm rad}}\text{,}
\end{equation}
where $\hbar\omega$ is the excitation energy (in our case, the ZPL energy), $n$ is the refractive index, $c$ is the speed of light, and $\varepsilon_0$ is the vacuum permittivity.

Substituting $E_{\rm ZPL} = 1.35\ \mathrm{eV}$, $n = 2.42$, and $|\mu| = 4.8\ \mathrm{D}$ into Eq.~\eqref{eq:tau}, we obtain $\Gamma_{\rm rad} = 2.25 \times 10^7\ \mathrm{s}^{-1}$, which gives a radiative lifetime of $\tau_{\rm rad} = 44.4\ \mathrm{ns}$ at zero pressure.

For context, the negatively charged SiV center exhibits a much shorter lifetime of approximately $1.7\ \mathrm{ns}$~\cite{Gali2013Theory, Zhang2018, Thiering2018, Zuber2023}, due to its larger transition energy. 

Under hydrostatic compression up to $\sim 117\ \mathrm{GPa}$, we observe that although the ZPL increases to $1.47\ \mathrm{eV}$ and the refractive index rises to $2.52$, the simultaneous reduction of the dipole moment to $2.95\ \mathrm{D}$ results in an increased lifetime of about $68.1\ \mathrm{ns}$. In contrast, under isotropic tensile strain of $-41.2\ \mathrm{GPa}$, the lifetime decreases slightly to $\sim 41\ \mathrm{ns}$. This decrease is associated with a ZPL energy of $1.30\ \mathrm{eV}$, a refractive index of $n = 2.40$, and an enhanced transition dipole moment of $|\mu| = 5.28\ \mathrm{D}$.

\section{Conclusions}\label{sec:conclusions}

We combined HSE06 first-principles calculations with a quadratic product Jahn--Teller (pJT) model to quantify the response of the neutral SiV center to extreme isotropic strain, spanning strong tensile loading to deep compression (from $\sim -80$ to $180$~GPa). In the symmetry-preserving $D_{3d}$ regime ($-41.2$ to $180$~GPa), the 946-nm zero-phonon line (ZPL) blue-shifts nearly linearly with pressure (overall slope $\sim0.8$--$0.9$~meV/GPa), providing a compact optical calibration for hydrostatic tuning. Throughout this $D_{3d}$ window, the ordering of the lowest vibronic excited states remains unchanged as the bright ${}^3\tilde{E}_u$ stays above the dark ${}^3\tilde{A}_{2u}$, and the vibronic gap $\delta$ increases under compression (to $\gtrsim8$~meV at $\sim180$~GPa) while it decreases toward the tensile stability limit. Hydrostatic strain preserves inversion symmetry, so parity stays well defined and the dark transition remains inactive.

A central result is that compression suppresses dynamical JT quenching by stiffening the relevant $E_g$ phonon and reducing the dominant stabilization energy $E_{\mathrm{JT}}^{(1)}$ (with the secondary branch $E_{\mathrm{JT}}^{(2)}$ driven toward zero), thereby increasing the Ham factor and substantially enhancing the observable excited-state spin--orbit splitting $\Delta_{\mathrm{SO}}$ (from $\sim25$~GHz near ambient to $\sim70$~GHz at the highest pressures considered). Tensile strain within the $D_{3d}$ window has the opposite effect, strengthening vibronic quenching and reducing $\Delta_{\mathrm{SO}}$.

Beyond a critical tensile strain ($\sim4\%$, $\sim-50$~GPa), relaxation yields two symmetry-equivalent $C_{3v}$ minima separated by a $D_{3d}$ saddle point, producing a double-well landscape. The corresponding tunneling-mediated interconversion ranges from GHz rates near the onset (effective dynamical averaging) down to MHz rates at larger tensile strain (static symmetry breaking). Charge-transition levels delineate practical operating limits: SiV remains photostable deep into compression, but the blue-shifting ZPL approaches an ionization threshold near $\sim100$~GPa; in the tensile symmetry-broken regime the ZPL jump renders the defect unstable under illumination. Finally, hyperfine couplings exhibit a systematic dependence on strain, decreasing under compression and increasing under tension. In contrast, the radiative lifetime is predicted to increase under strong compression, as the rapid decrease in the transition dipole moment compensates for the opposing cubic scaling of the increasing ZPL energy. Together, the ZPL position, spin-orbit splitting, and selected hyperfine parameters provide a compact multi-observable calibration of isotropic strain, establishing SiV as a symmetry-protected, strain-tunable quantum emitter into the multi-megabar-equivalent regime.

\section*{Acknowledgement}

We acknowledge the discussion with Zoltán Sántha. This work was supported by the Hungarian National Research,
Development and Innovation Office (NKFIH) for Quantum Information National
Laboratory of Hungary (grant no.\ 2022-2.1.1-NL-2022-00004), the EU QuantERA II
SensExtreme project funded by NKFIH (grant no.\
2019-2.1.7-ERA-NET-2022-00040) and the Horizon Europe Quantum Flagship SPINUS project
(grant no.\ 101135699). G.\ T.\ was supported by the J\'anos Bolyai Research
Scholarship of the Hungarian Academy of Sciences and by NKFIH grant no.\
STARTING 150113.
This work was conducted by Meysam Mohseni under the supervision of Gerg\H{o} Thiering and Adam Gali. 

\bibliography{SiC}
\end{document}